\documentclass{desyproc}

\begin{document}
\title{Gluon saturation effects at forward rapidities at LHC in pp collisions}

\author{{\slshape Amir H. Rezaeian }\\[1ex]
Institut f\"ur Theoretische Physik, Universit\"at Regensburg, 93040 Regensburg, Germany\\
Departamento de F\'\i sica y Centro de Estudios Subat\'omicos,\\ Universidad T\'ecnica Federico Santa Mar\'\i a, Casilla 110-V, Valpara\'\i so, Chile }

\contribID{xy}

\confID{800}  
\desyproc{DESY-PROC-2009-xx}
\acronym{LP09} 
\doi  

\maketitle

\begin{abstract}

We investigate hadrons and direct photon production in pp collisions
at the LHC energy within the color-dipole approach. We show that
greatest sensitivity to gluon saturation effects is reached at very
forward rapidities in pp collisions at LHC ($\sqrt{s}=14~\text{TeV}$).  The
discrepancies among various saturation models (fitted to HERA data) results can
be about a factor of $2-3$ at forward rapidities.  We found that the
ratio of direct-photon to pion production can be about $20-10$ at
forward rapidities $\eta=7-8$. Therefore, direct photon production at forward rapidities
should provide a rather clean probe as the background from
radiative hadronic decays is significantly
suppressed. 

\end{abstract}

\section{Introduction}
At high energies/small Bjorken-x, QCD predicts that gluons in a hadron wavefunction form a new state, the so-called
Color Glass Condensate (CGC) \cite{Gribov:1984tu1,ocgc}, for a review see
\cite{Iancu:2003xm} and references therein.
The cornerstone of the CGC is the existence of a hard saturation
scale $Q_s$ at which nonlinear gluon recombination effects become important and start to
balance gluon radiation. The Color Glass Condensate (saturation) approach to QCD at high energy
has been very successful to describe a variety of processes at
Relativistic Heavy Ion Collider (RHIC) \cite{Iancu:2003xm}. Nevertheless, the
importance of saturation effects is still disputable given that other
approaches offered alternative descriptions. In order to test
saturation physics and its relevance, it seems therefore essential to
consider various reactions in different kinematic regions at the Large Hadron Collider (LHC) and
future collider experiments.  Here, we address the role of gluon saturation at LHC energy $\sqrt{s}=14$ TeV in hadrons and direct photon production 
in $pp$ collisions within the light-cone color-dipole approach using various saturation models. Details of calculations can be found in Ref.~\cite{RS}.
 
\section{Hadrons and photon at LHC within saturation models }
\begin{figure}[!t]
       \includegraphics[width=6 cm] {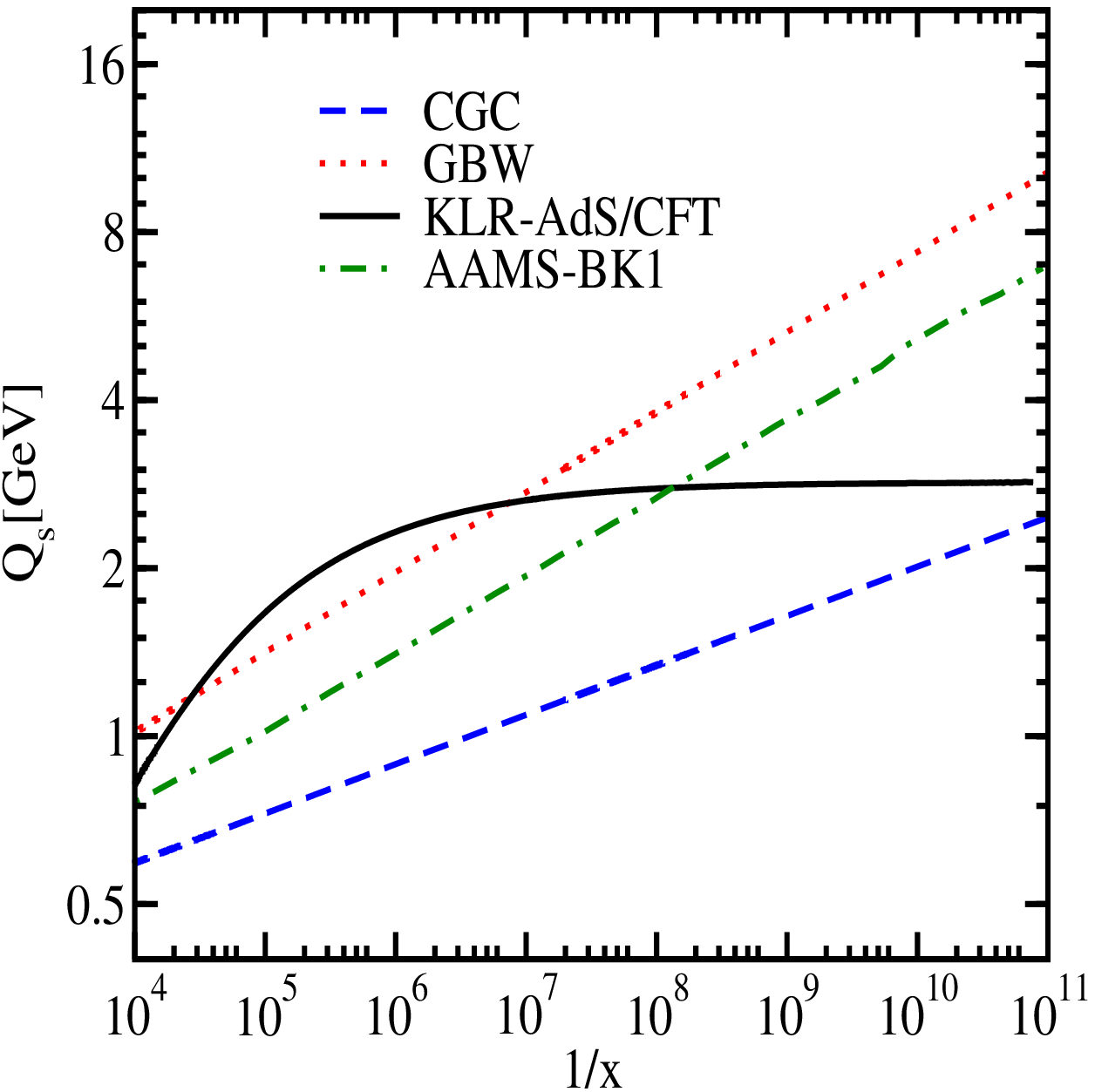}
       \includegraphics[width=6 cm] {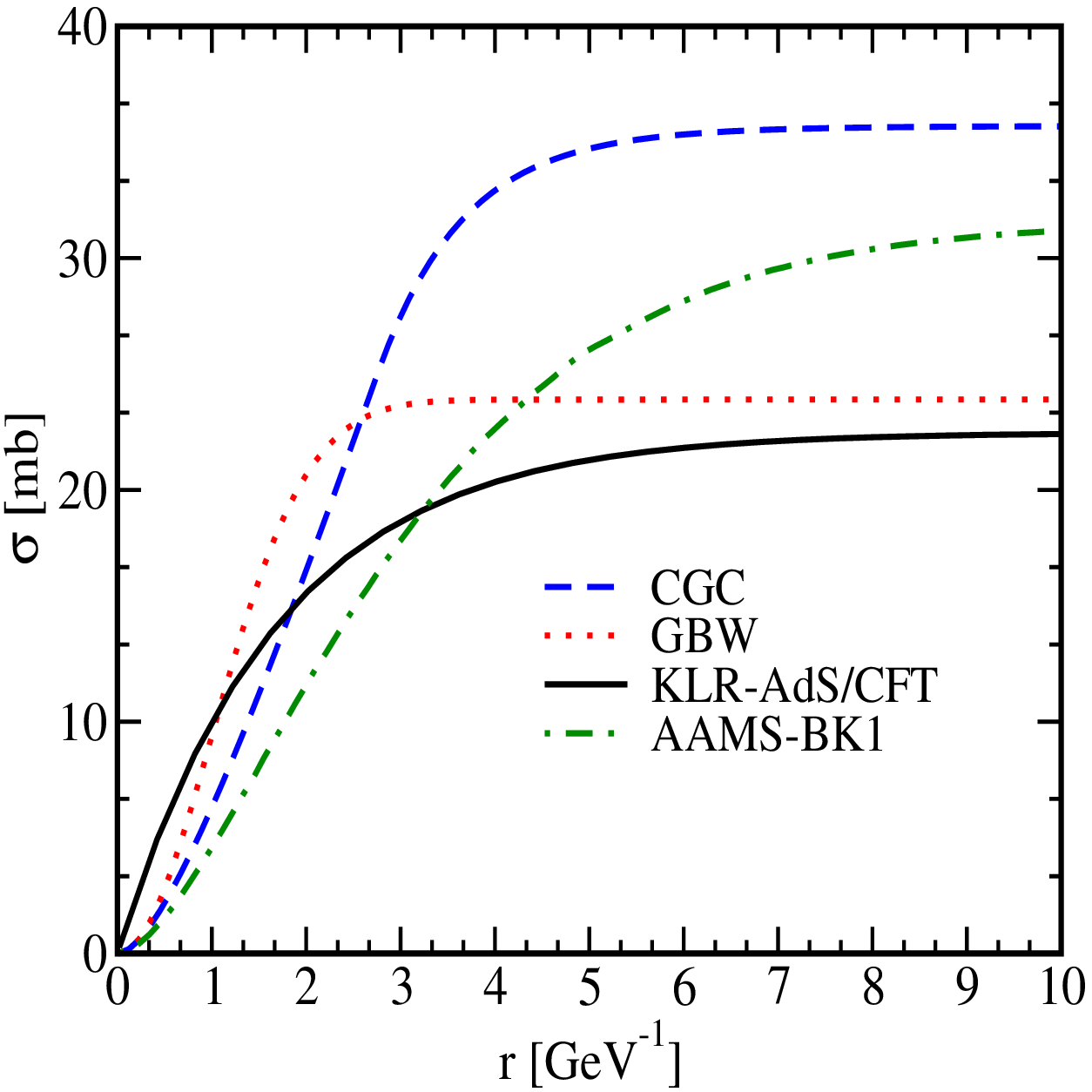} 
       \caption{Left: Saturation scale as a function of $1/x$ for various
       color-dipole models labeled with CGC \cite{cgc}, GBW \cite{gbw}, KLR-AdS/CFT \cite{ads} and AAMS-BK1 \cite{aams}.  Right: The total dipole-proton cross
       section $ \sigma_{q\bar{q}}(r,x)$ at fixed $x=10^{-5}$ in the various
       color-dipole models. } \label{f:sat}
\end{figure}

The concept of saturation and the taming of the power-like rise of the
gluon distribution at small Bjorken-x was first addressed in Ref.~\cite{Gribov:1984tu1} in
the double logarithmic approximation. The actual calculation of higher-order corrections to 
the non-linear small-x evolution equations still remains as a challenge \cite{Iancu:2003xm}, see also \cite{ads} and references therein. 
Thus, one may resort to a QCD-like model which incorporates the basic
features of gluon saturation into the dipole-proton forward scattering
amplitude, and provides predictions which will allow to test the validity of the treatment. There are several parametrizations
proposed in the literature which all give a good description of HERA
data but predict different saturation scales, see Fig.~\ref{f:sat}.
The details of saturation models used in Fig.~\ref{f:sat} can be found in Ref.~\cite{RS}.
The main feature of these models is that for decreasing $x$, the dipole
amplitude saturates at smaller dipole sizes.  Note that there is no
unique definition for the saturation scale in literature.  We define the saturation scale
$Q_s^2=2/r_s^2$ as a energy scale at which the $q\bar{q}$ dipole
scattering amplitude $\mathcal{N}_{q\bar{q}}\approx 0.4$ becomes sizable \cite{RS,gbw}.

\begin{figure}[!t]
       \includegraphics[width=6 cm] {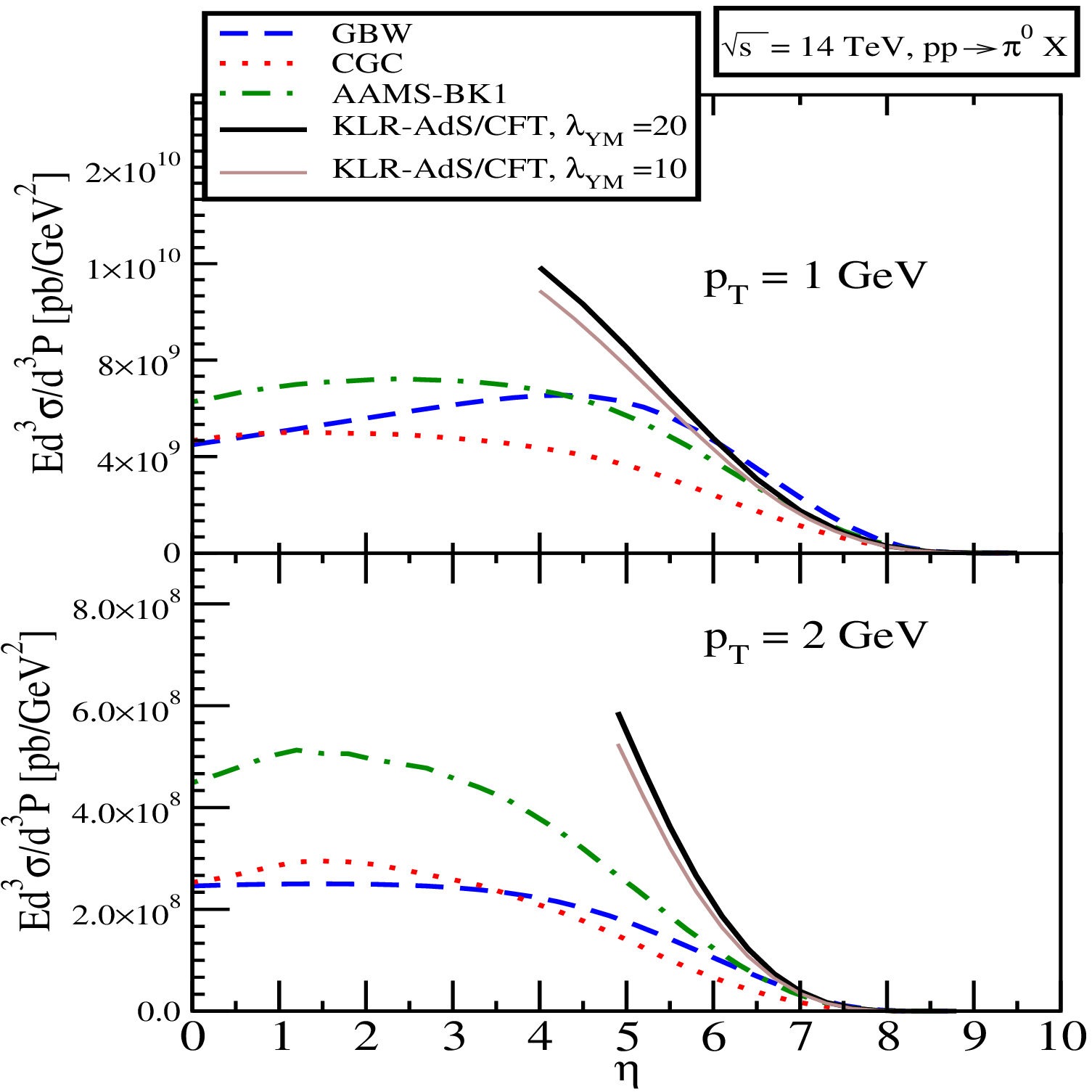}
       \includegraphics[width=6 cm] {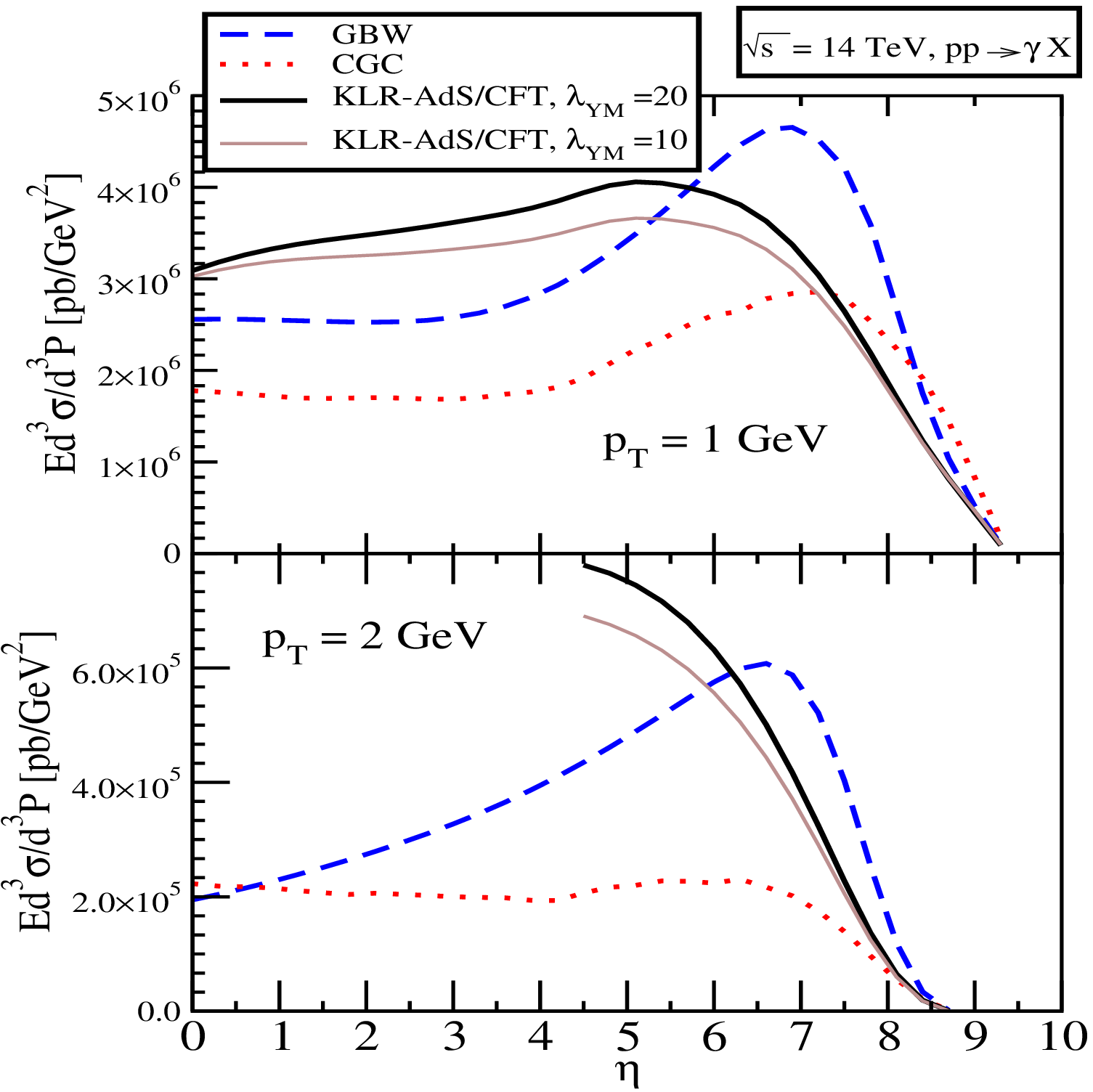}
        \caption{Invariant cross-section for pion (left) and direct photon (right panel) production in $pp$ collisions at LHC as a
function of rapidity $\eta$ calculated with various color dipole models for various fixed
$p_T$. \label{lhc-e1}}
\end{figure}

The invariant cross-section of hadron and direct photon production can be calculated via the light-cone color-dipole
factorization scheme \cite{RS}, see also Ref.~\cite{me1}. In Fig.~\ref{lhc-e1}, the differential cross-section of pion $\pi^0$
and direct photon $\gamma$ (photons radiated in hadronic collisions not via
hadronic decays) production at LHC are plotted versus rapidity at
fixed transverse momenta $p_T =1$ and $2$ GeV within various
color-dipole models (see Fig.~\ref{f:sat}). It is seen that the discrepancies among various
saturation color dipole model results can be about a factor of $2-3$ at
moderate rapidities.  At the kinematic limit, i.e. at very forward
rapidities and higher $p_T$ where the differential cross-section
approaches zero, kinematic constraints limit the parton phase space
and saturation effects become less important. It is seen from Fig.~\ref{lhc-e1} that for both hadron and
photon production, away from the kinematic limit, at not very large $\eta$ and
$p_T$, a color-dipole model with larger saturation scale leads to a stronger 
peak at forward rapidity (having in mind that the saturation scale is a dynamical function of $x$, see Fig.~\ref{f:sat}).

\begin{figure}[!t]
       \includegraphics[width=6 cm] {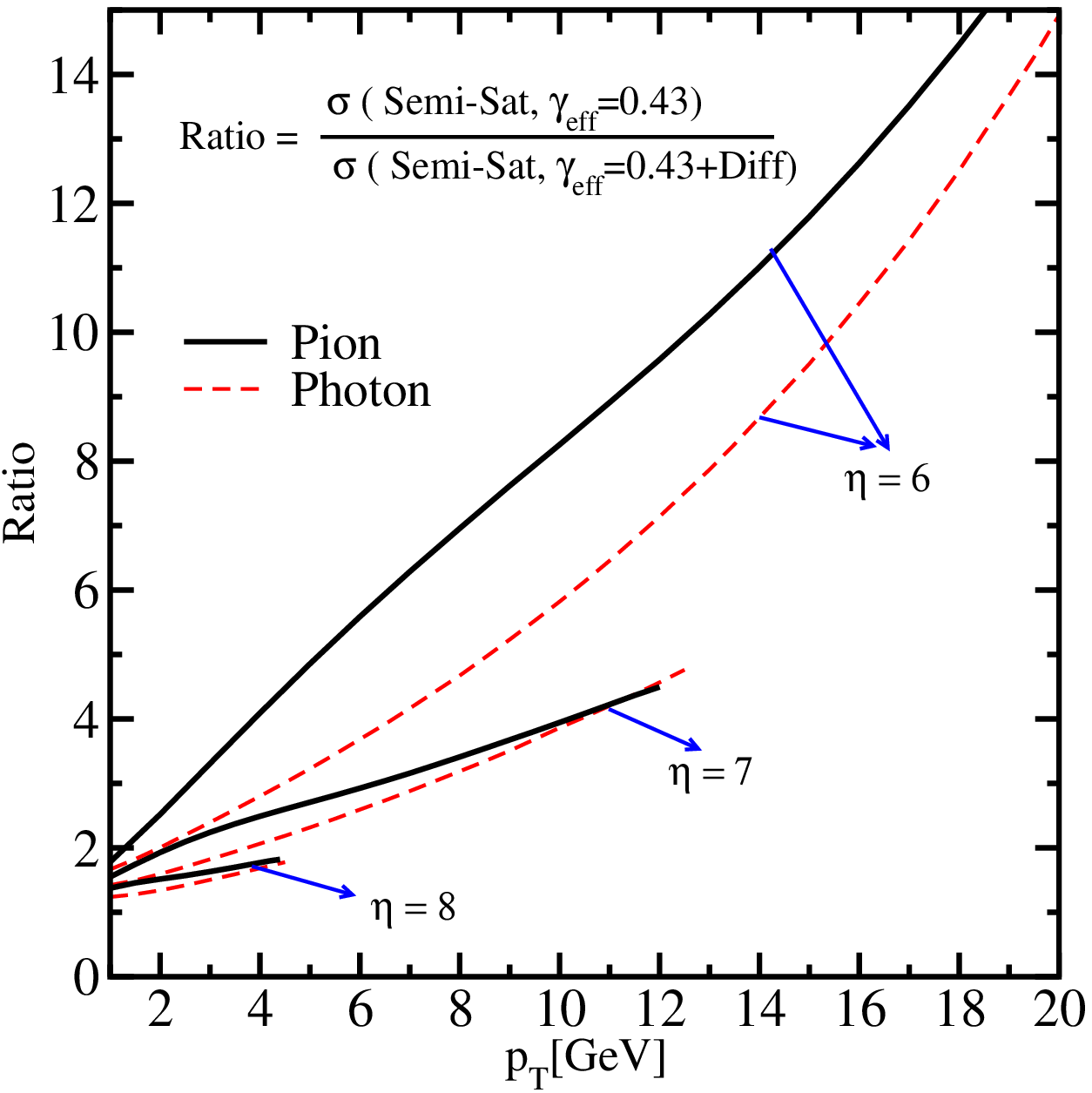}
       \includegraphics[width=6 cm] {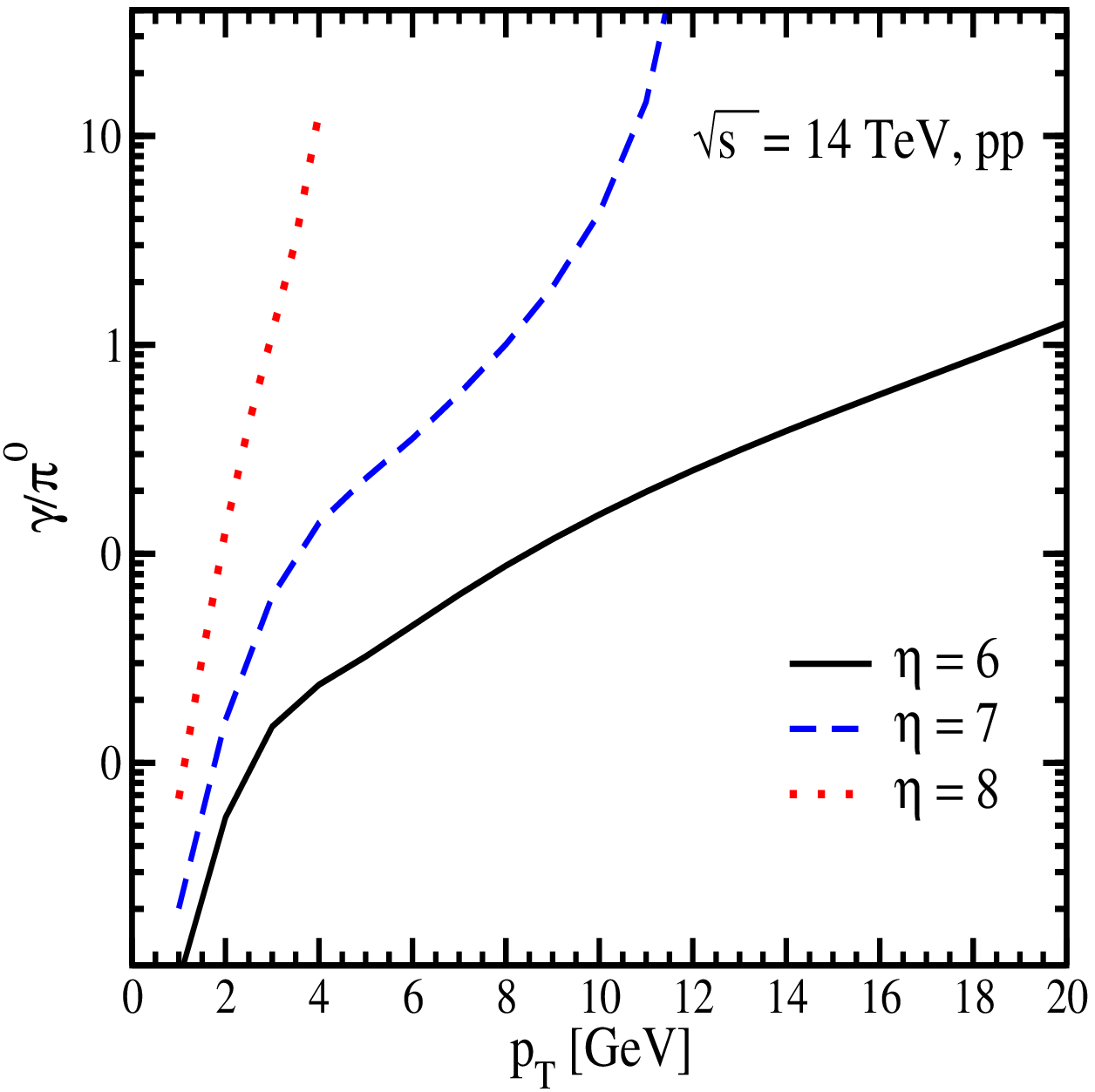} \caption{Left: The ratio of two invariant cross-sections obtained
       by taking two different effective anomalous dimension
       $\gamma_{eff}$ in the Semi-Sat dipole model for pion and direct
       photon production at LHC energy ($\sqrt{s}=14$ TeV) in $pp$
       collisions at various rapidities. Right:
       The ratio of photon/pion production in $pp$ collisions at LHC
       as a function of $p_T$ calculated with the GBW color dipole
       model. \label{lhc-s1} }
\end{figure}

In order to understand the {\em relative} importance of saturation effects
at various rapidities, we employ the Semi-Sat model fitted to HERA data with the dipole-proton forward scattering
amplitude  \cite{bccc}: 
\begin{equation} 
\mathcal{N}_{q\bar{q}}^{\text{Semi-Sat}} (\vec{r}, \vec{b}, x)=2\mathcal{N}_0\left(\frac{rQ_s}{2}\right)^{2\gamma_{eff}}, \hspace{2cm} Q_s=\left(\frac{x_0}{x}\right)^{\frac{\lambda}{2}}\;\left[\exp\left(-\frac{b^2}{2B_{\rm CGC}}\right)\right]^{\frac{1}{2\gamma_s}}. \label{nos}
\end{equation}
 The parameter $\gamma_{eff}$ is defined for $rQ_s\le 2$ as $\gamma_{eff}=\gamma_s +\frac{1}{\kappa\lambda Y}\ln\frac{2}{rQ_s}$,
and for $rQ_s> 2$ as $\gamma_{eff}=\gamma_s$. The value of other parameters of this model can be found in \cite{bccc}.  Surprisingly, the fit obtained with such
an oversimplified model is as good as for the other models with $\chi^2/\text{d.o.f.}=0.92$. 
In Fig.~\ref{lhc-s1} we show, ratio of the two cross-sections for both pions and direct photon at LHC, calculated once
with diffusion term and once without, i.e. $\gamma_{eff}=0.43$.
It is seen that at forward rapidities, the diffusion term in the anomalous dimension is
not important, since it gives similar results as with a fixed
$\gamma_{eff}=0.43$. The preferred value of anomalous dimension $1-\gamma_{eff}=0.57$ at very forward rapidities
is close to the one predicted from the BK equation \cite{bkk}.  
It is well known that the saturation effects start being
essential when the anomalous dimension reaches the value $\gamma_{cr}
= 1-\gamma_{eff}=0.37$ which is the case for forward rapidities (see Refs.~\cite{Gribov:1984tu1,lbk}). This
indicates that direct photon and hadron production at different
rapidities at LHC are rather sensitive to saturation. 

Direct photons can only be 
radiated from quarks, while hadrons can be produced by both gluons and
quarks. At the LHC energy at midrapidity gluons dominate. Therefore
the photon/pion ratio is significantly reduced toward midrapidity. However, at
very forward rapidity, valence quarks become important and the
photon/pion ratio rises. Moreover, at high $p_T$ again valence quarks
becomes important and we have a sharp rise of the photon/pion ratio.
In Fig.~\ref{lhc-s1} right panel, we
show the photon/pion ratio $\gamma/\pi^0$ as a function of $p_T$ at
various rapidities within the GBW model and $pp$ collisions. The ratio
$\gamma/\pi^0$ can be as big as $10-20$ at very forward
rapidities $\eta=8-7$ at LHC energy. Therefore, direct photon production extends to higher
rapidities for a fixed $p_T$, see Figs.~\ref{lhc-s1}. Note that
suppression of hadrons at very forward rapidity also ensures significant
suppression of radiative decays of those hadrons. Therefore, measurements of direct photons at
forward rapidities should be rather clean, as the background from
radiative hadronic decays is significantly
suppressed.

\section{Acknowledgments}
A.R. acknowledges the financial support from the Alexander von Humboldt foundation, BMBF (Germany), Conicyt Programa
  Bicentenario PSD-91-2006, Fondecyt grants 1090312 (Chile).


\begin{footnotesize}


\end{footnotesize}


\end{document}